\documentclass{article}
\usepackage{ijcai23}

\usepackage{times}
\usepackage{soul}
\usepackage{url}
\usepackage{pgfplots}
\usepackage[hidelinks]{hyperref}
\usepackage[utf8]{inputenc}
\usepackage[small]{caption}
\usepackage{graphicx}
\usepackage{amsmath}
\usepackage{amsthm}
\usepackage{booktabs}
\usepackage{algorithm}
\usepackage{algorithmic}
\usepackage[switch]{lineno}
\usepackage{mathtools}
\usepackage{amssymb}
\usepackage{xcolor}
\usepackage{bm}
\usepackage{enumitem}
\usepackage[flushleft]{threeparttable}
\usepackage{diagbox}
\usepackage{multirow}

\theoremstyle{plain}

\theoremstyle{definition}

\theoremstyle{remark}

\DeclareMathOperator*{\argmin}{arg\,min}
\DeclareMathOperator*{\argmax}{arg\,max}
\newcommand{\y}{\bm{y}}

\newcommand{\x}{\bm{x}}
\newcommand{\e}{\mathrm{e}}
\newcommand{\norm}[1]{\lvert \lvert #1 \rvert\rvert}
\newcommand{\w}{\bm{w}}
\newcommand{\R}{\mathbb{R}}

\newcommand{\p}{\bm{p}}
\renewcommand{\t}{\bm{\theta}}
\renewcommand{\o}{\bm{\omega}}
\renewcommand{\u}{\bm{u}}
\renewcommand{\sp}{\vspace{3pt}}

\newcommand{\pslpred}[1]{\textsc{#1}}
\newcommand{\pslarg}[1]{\texttt{#1}}
\newcommand{\psland}{\wedge}

\newcommand{\pslneg}{\neg}

\newcommand{\pslthen}{\rightarrow}

\title{DeepPSL: End-to-end perception and reasoning}

\author{
Sridhar Dasaratha$^1$\and
Sai Akhil Puranam$^1$\and
Karmvir Singh Phogat$^1$\and
Sunil Reddy Tiyyagura$^1$\and
Nigel Duffy$^2$
\affiliations
$^1$ EY Global Delivery Services India LLP\\
$^2$ Ernst \& Young (EY) LLP USA 
\emails
\texttt{\{Sridhar.Dasaratha,Sai.Puranam,Karmvir.Phogat\}@gds.ey.com},
\texttt{Sunil.Tiyyagura@gds.ey.com},
\texttt{Nigel.P.Duffy@ey.com}
}

\begin{document} 
\maketitle
\begin{abstract}
We introduce DeepPSL a variant of probabilistic soft logic (PSL) to produce an end-to-end trainable system that integrates reasoning and perception. PSL represents first-order logic in terms of a convex graphical model -- hinge-loss Markov random fields (HL-MRFs). PSL stands out among probabilistic logic frameworks due to its tractability having been applied to systems of more than 1 billion ground rules. The key to our approach is to represent predicates in first-order logic using deep neural networks and then to approximately back-propagate through the HL-MRF and thus train every aspect of the first-order system being represented. We believe that this approach represents an interesting direction for the integration of deep learning and reasoning techniques with applications to knowledge base learning, multi-task learning, and explainability. Evaluation on three different tasks demonstrates that DeepPSL significantly outperforms state-of-the-art neuro-symbolic methods on scalability while achieving comparable or better accuracy.
\end{abstract}

\section{Introduction}

Many machine learning problems involve rich and structured domains with numerous dependencies between its elements. Statistical relational learning (SRL) \cite{MLN,koller2007graphical} methods seek to represent these dependencies and create graphical models using rule based representations. A fundamental challenge faced by SRL approaches is balancing scalability with the expressivity of the dependency structure.

\cite{bach2017hinge} introduced HL-MRFs a class of probabilistic graphical models that are both tractable and expressive enabling scalable modelling of rich structured data. In addition they provide a powerful  formalism, probabilistic soft logic (PSL) that can define the HL-MRF using first order logic and introduce a scalable inference algorithm. The continuous nature of HL-MRFs enable PSL to scale beyond what was previously feasible for SRL frameworks \cite{trivialblocking}. Using PSL, problems with tens of millions of ground rules have been solved in minutes \cite{kouki2017collective}. Recent advances using tandem inference make inference tractable even for extremely large systems (billions of random variables) \cite{srini20}. The PSL technique has been successfully applied to problems from various domains ranging from knowledge extraction \cite{rospocher2018ontology}, cyberattack prediction \cite{perera2018cyberattack}, relational fairness \cite{fairness18}, enrichment of product graphs \cite{gandoura2020human} to hybrid recommender systems \cite{rodden2020vmi}.  

While PSL has significantly advanced SRL methods, there have also been remarkable advances in the field of perception driven by deep learning methods. It would be highly desirable to integrate these perception capabilities into the PSL framework: however currently there is no mechanism to provide this integration. We tackle this challenge with DeepPSL, an end-to-end  integration of PSL with deep learning thus achieving a major enhancement of PSL capabilities. DeepPSL fully inherits the scalability of PSL both during inference and training.

The first order expressions in PSL are built from predicates that capture the truth of an assertion with soft truth values in [0,1]. For instance, “HasClaws” and “HasStripes” could be predicates that represent whether claws and stripes are detected in an image. Given some mechanism to compute these truth values and combined with knowledge of animal attributes, PSL can infer  whether a specific animal is present in the image. On the other hand neural nets (NN) can learn to identify animals directly from an image, typically from large quantities of training data. 

With DeepPSL we integrate these two approaches: some of the predicates are replaced by neural nets and the input data (for e.g., text or image) is processed through a NN to generate the predicate values which are then used by PSL for inference. End-to-end training of this architecture on training data for a given task then permits the NN to learn concepts without any data on the concept itself. In contrast to PSL where one must define the predicate, the training in DeepPSL directly learns predicates that are optimized for the task at hand. Further, these concepts can now be utilized in other tasks where we may have only limited or no task specific data. 

Training of this architecture poses significant challenges as it requires back-propagation through an HL-MRF that does not have continuous derivatives. This optimization problem cannot be solved by adapting existing convex optimization methods but solving it is critical to integrating deep learning and PSL. The key contribution of our work is a novel and non-obvious approach to back-propagate through the HL-MRF to learn the parameters of these deep networks. The proposed algorithm enables end-to-end training of DeepPSL which in turn helps fully realize the benefits of the proposed architecture. 

We evaluate the efficacy and performance of DeepPSL on three different tasks: digit addition, semi-supervised classification and entity resolution. Experiment results demonstrate the superior scalability of DeepPSL over other state-of-the-art neuro-symbolic approaches while achieving similar or better accuracy.  

\section{Related Work}
Relational Neural machine (RNM) \cite{marrarelational}, an extension of Deep logic models (DLM) \cite{Deep_Logic_Models}, model reasoning using a Markov random field and backpropagate through that field to learn underlying neural models. Unlike DeepPSL, RNMs do not require backpropagation through an $\argmin$ and do not allow any learned values to be used directly in logical rules. Rather they add potentials that couple the learned values to output variables which must be either observed or latent in the Markov random field potentially resulting in a large increase in the number of latent variables. DLM and RNM are related to semantic-based regularization \cite{semantic_regularization}, logic tensor networks \cite{Logic_Tensor_Networks} and neural logic machines \cite{Neural_Logic_Machines} which allow logical constraints to constrain the learning of deep networks.

Neural theorem prover \cite{neural_theorem_provers} is an end- to-end differentiable prover. TensorLog \cite{TensorLog} is a recent framework to reuse the deep learning infrastructure of TensorFlow to perform probabilistic logical reasoning. Neither of these methods model predicates using deep learning. \cite{Logic_DL} presents an iterative distillation method that transfers structured information of first-order logic rules into the weights of the NNs. \cite{PSL_SL_Neural} generalized the approach to include rules built using PSL.

DeepProbLog \cite{DeepProbLog} augments the probabilistic logic programming language ProbLog \cite{ProbLog} by incorporating neural predicates, and makes predictions by employing marginal inference and sampling. DeepProbLog and other methods such as NeurASP \cite{neur_asp} do not scale well as they rely on the computationally expensive possible world semantics. DeepStochLog \cite{winters2021deepstochlog} achieves better scalability as compared to DeepProbLog and NeurASP using stochastic definite clause grammars. In contrast, DeepPSL scales to significantly larger systems with millions of ground rules by leveraging the continuous nature of HL-MRFs that cast MAP inference as a convex optimization problem. \cite{neu_psl} propose a neuro-symbolic approach based on an energy based modeling extension of the PSL framework. Their work uses a novel "energy loss" that doesn't require back propagating through a convex optimization problem. Our approach allows for arbitrary convex differentiable loss functions including all of the most commonly used losses. 

End-to-end training of a DeepPSL model requires solving a bi-level optimization problem. The techniques discussed in \cite{sinha2017review,ghadimi2018approximation,dempe2018bilevel} for solving a bi-level optimization computes gradient of the loss function which needs computation of inverse of the Hessian that is expensive to compute at each iteration. Other methods consider neural network layers consisting of a variety of optimization problems: $\argmin$ and $\argmax$ problems \cite{GouldFernando}, quadratic programming problems \cite{AmosKolter,LeeMajiRavichandran}, convex problems \cite{AgarwalAmosBarratt}, cone programs \cite{AgarwalBarrattBoyd}. All of these methods require that the optimization functions have continuous derivatives and make use of second derivatives to allow backpropagation through the optimization problems. HL-MRFs do not have continuous derivatives and therefore are not amenable to these approaches.
  
\section{Background}
\subsection{HL-MRFs: Hinge Loss Markov Random Fields}\label{ssec:hlmrf}
HL-MRFs are defined with $k$ continuous potentials $ \bm{\phi} =\{\phi_1,\dots,\phi_k\} $ of the form:
\begin{equation}\label{eq:potential}
\phi_j(\x, \y) = (max\{l_j(\x, \y),0\})^{p_j}
\end{equation}
where $\phi_j$ is a potential function of $n$ free random variables  $\y = \{y_1,\dots,y_n\}$  conditioned on $n'$ observed random variables $\x = \{x_1,\dots,x_{n^{'}}\}$, each random variable can take soft values between $[0, 1]$. The function $l_j$ is linear in $\y$ and $\x$ and $p_j \in \{1,2\}$.
\footnote{In this work, $p_j = 2$ is used for quadratic hinge-loss.}
Collecting the definitions from above, a hinge-loss energy function $f$ is defined as
\begin{equation}\label{energy_function}
f_{\t}(\x, \y) = \sum\limits_{j=1}^{k} \theta_j \phi_j(\x, \y)
\end{equation}
where $\t= \left(\theta_1, \ldots, \theta_k\right)$ and $\theta_j$ is a positive weight corresponding to the potential function $\phi_j$. 
A HL-MRF over random variables $\y$ and conditioned on random variables
 $\x$ is a probability density defined as
\begin{equation}\label{equation_pd}
P(\y|\x) = \frac{1}{Z(\x)} \exp{(- f_{\t}(\x, \y) )}
\end{equation}
where $Z(\x)$ is the partition co-efficient. Maximum a posteriori (MAP) inference finds the most probable assignment to the free variables $\y$ given the observed variables $\x$. MAP inference is done by maximizing the probability density $P(\y|\x)$ while satisfying the constraint that the random variable $\y \in [0, 1]^{n}$. Since the normalizing function $Z$  in \eqref{equation_pd} is not a function of $\y$, maximizing $P(\y|\x)$ is equivalent to minimizing the
energy function $f$, i.e., 
\begin{equation} \label{eq:psl}
\underset{\y \in [0, 1]^n }{\arg\max} \, P(\y|\x) \equiv  \underset{\y \in [0, 1]^n }{\argmin}\,f_{\t}(\x, \y)
\end{equation}
Critically the function $f$ is convex in $\y$, for a given $\x$, allowing for tractable inference even for very large HL-MRFs. In this study, the inference problem is solved by employing stochastic gradient descent\footnote{The SGD optimization in PSL suffers from a drawback that hard constraints cannot be strictly enforced. However, this limitation can be circumvented by employing ADMM in place of SGD.} (SGD) algorithm.  However, one may employ other alternative algorithms such as distributed optimization algorithm, alternating direction method of multipliers (ADMM), as discussed in \cite{bach2017hinge}.

\subsection{PSL rules}
PSL uses first order logic as a template language for HL-MRFs. A PSL program defines a set of rules in first order logic. These rules in the PSL program are grounded over the base of ground atoms each of which represents an observation or unknown of interest. These ground atoms are associated with random variables $\left(\x, \y\right)$ and can take any value in $[0,1]$. Each ground rule is then translated into a weighted hinge-loss potential. The sum of these potentials defines a HL-MRF, and minimizing the HL-MRF conditioned over $\x$ gives values for the inferred predicates $\y$.

It is beyond the scope of this paper to provide a detailed description of how first order logic rules are used as a template language for HL-MRFs, see \cite{bach2017hinge} for further details.

\section{DeepPSL}
\subsection{Deep Learning based Predicates}
The key difference between PSL and DeepPSL is that some of the predicates are modeled with deep neural networks  (DNNs). In PSL, the observed predicates $\x$ are available through a knowledge base, while in DeepPSL a feature vector $\u$ is processed through DNNs to compute some of the predicates. Typically, there is no data available on these predicates to learn their corresponding DNNs. Hence, end-to-end training of DeepPSL system is needed to learn these predicates.

\subsection{Learning}
The key problem that needs to be solved is to determine how to train this system end-to-end. In the proposed DeepPSL framework, the features $\u$ are first processed through a neural net with tunable weights $\o$ to generate estimates of $\x$ which are predicates for the PSL. The estimates of predicates in the DeepPSL are modeled by a deep neural network $\p(\u;\o)$. These predicates then go through PSL inference to produce the final values of the random variables $\y$. For end-to-end training, we need to enable backpropagation through the PSL inference. Since PSL inference is a convex optimization problem, there is no direct way to backpropagate and update the weights of the predicate network. We now describe our solution to address this problem.

\subsubsection{Optimization objective}
The prime objective of training this end-to-end learning model is to determine weights $\w = (\o, \t)$ such that the HL-MRFs inference yields variables $\y$ which are close to their true values $\hat{\y}$ in the training data. These free variables $\y$ represent the outputs of DeepPSL. For example, a $\y$ might represent a belief that a given image contains a zebra. 

We want to obtain good outputs or predictions where ``good" is measured by a loss relative to their true values $\hat{\y}$. We restrict our analysis here to HL-MRFs in which all $\y$ correspond to outputs.

In order to measure if the inferred values $\y$ are close enough to the true values $\hat{\y}$; let us consider a differentiable convex loss function: 
\begin{equation}\label{eqn_cel}
\R^{n} \times \R^{n} \ni (\hat{\y} , \y ) \mapsto L(\hat{\y}, \y)  \in \R
\end{equation}
The DeepPSL inference problem \eqref{eq:psl} is approximated with soft constraints \footnote{The constraints in \eqref{eq:psl} are incorporated using Lagrange multipliers in a fairly standard way.} and $\w = \left(\o, \t\right)$ as
\begin{equation}\label{eq:a_psl}
\y^* = \underset{\y}{\argmin} \;\;\tilde{f} (\u, \w, \y)
\end{equation}
where 
\begin{align*}
\tilde{f}(\u, \w,\y) = & f_{\t} (\p(\u;\o), \y) +  \sum_{i=1}^{n} \underline{\gamma_i} (\max\{0, -y_i\})^2 \\
& + \sum_{i=1}^{n} \overline{\gamma_i} (\max\{0, y_i - 1\})^2
\end{align*}
with fixed $\overline{\gamma_i}, \underline{\gamma_i} > 0$.
Therefore, the weight training problem is set up as a nonlinear optimization
\begin{equation}\label{eq:opt}
\begin{aligned}
\min_{\w,\y} &\;\; L(\hat{\y}, \y) \\
\text{subject to } &\;\; \y = \underset{\bar{\y}}{\argmin} \;\;\tilde{f} (\u, \w, \bar{\y}) 
\end{aligned}
\end{equation}

\subsubsection{Gradient Following Algorithm}
We develop a gradient descent procedure for solving the nonlinear optimization \eqref{eq:opt}.
This task is challenging because we need to back-propagate through the $\argmin$.
The most direct approach involves inverting the Hessian $\nabla_{\y\y} \tilde{f} (\u, \w, \y)$ which is not well-defined for HL-MRFs as they do not have continuous derivatives.

We take an alternative approach which avoids this pitfall. We assume that the functions $L$ and $f$ are differentiable. Furthermore, we assume that the gradient $\nabla_{\y} \tilde{f}$ and the neural network $\p$ are locally Lipschitz continuous. In general, DNNs are designed to be trained using gradient based techniques and that requires DNNs to be locally Lipschitz continuous, see \cite{roc81,scaman2018lipschitz,jordan20}. Therefore, these assumptions are general enough and the proposed technique is applicable to a wide class of problems. 

Consider the neural network weights $\w_{t-1}$ such that $\y_t = \argmin_{\y} \tilde{f}(\u,\w_{t-1},\y)$. The objective is to find a $\w_t$ such that
\begin{align*}
L(\hat{\y}, \y_{t+1}) < L(\hat{\y}, \y_t) \;\; \text{where} \;\; \y_{t+1} = \argmin_{\y} \tilde{f}(\u,\w_{t},\y)
\end{align*}
To this end, we first linearly approximate the constraint $\tilde{\y}_{t+1} = \argmin_{\y} \tilde{f}(\u,\w,\y)$ in the neighborhood of $\y_t$ by using continuous dependence\footnote{Please note that, in general,  convexity of a differentiable function $f$ is not sufficient to ensure continuous dependence of $\argmin_{\y} \tilde{f}(\u, \w, \y)$ on $\w$. However, $\argmin_{\y} \tilde{f}(\u, \w, \y) + \nu \norm{\y - \y_t}^2 $ for any $\nu > 0$ is augmented to ensure uniqueness of the solution and so continuous dependence on $\w$.} of $\tilde{\y}_{t+1}$ on $\w$ and that is given by   
\begin{align}\label{eq:app_y}
\tilde{\y}_{t+1} = \y_t - \delta \nabla_{\y} \tilde{f}(\u,\w,\y_t) 
\end{align}
for sufficiently small $\delta > 0.$
Therefore, using the approximation \eqref{eq:app_y}, the optimization \eqref{eq:opt}, in the neighborhood of $\y_t$, translates to 
\begin{align}\label{eq:app_opt}
\argmin_{\w} L(\hat{\y}, \y_t - \delta \nabla_{\y} \tilde{f}(\u,\w,\y_t))
\end{align} 
and that is linearly approximated, in the neighborhood of $\y_t$, to 
\begin{equation}\label{eq:w_t}
\w_t = \argmax_{\w} \nabla_{\y} \tilde{f}(\u, \w, \y_t) \cdot \nabla_{\y} L(\hat{\y}, \y_t) 
\end{equation}
It is worth noting that if $\nabla_{\y} \tilde{f}(\u, \w_t, \y_t) \cdot \nabla_{\y} L(\hat{\y}, \y_t) > 0$ then $L(\hat{\y},\tilde{\y}_{t+1}) < L(\hat{\y},\y_t).$ On the other hand if $\nabla_{\y} \tilde{f}(\u, \w_t, \y_t) \cdot \nabla_{\y} L(\hat{\y}, \y_t) \leq 0$ then local optimality is attained at $\w_{t-1}$. Furthermore, to ensure constraint satisfaction at each iteration, the local linear approximation $\tilde{\y}_{t+1}$ is replaced with the inference optimization 
\begin{align}\label{eq:inf}
\y_{t+1} = \argmin_{\y} \tilde{f}(\u,\w_t,\y)  
\end{align}
Recall that $\lim_{h \rightarrow 0} \frac{g(v+hz) - g(v)}{h} = \nabla g(v) \cdot z$ and therefore, \eqref{eq:w_t} can be rewritten as  
\begin{equation}\label{eq:convex_difference}
\w_t = \argmin_{\w} \; \tilde{f}(\u, \w, \y_t - \alpha \nabla_{\y}L(\hat{\y},\y_t)) - \tilde{f}(\u,\w, \y_t)
\end{equation}
for sufficiently small $\alpha>0$.

\subsubsection{Regularization}\label{sssec:dying-hinge-loss}
A PSL potential $\phi_j$ corresponding to a rule $j$, defined in \eqref{eq:potential}, translates in DeepPSL setup to 
\[
\phi_{j}(\u, \o, \y) = \max \left(l_{j}(\p(\o; \u),\y), 0\right)^{p_j} 
\] 
where ${l}_{j}$ is a linear function, $p_j \in \{1,2\}$, $\p(\o;\u)$ is a neural network with weights $\o$ and input $\u$. Note that, for a certain $\tilde{\o}$, the potential $\phi_j$ does not trigger, i.e.,  
\[
\phi_{j}(\bar{\u}, \tilde{\o}, \bar{\y}) = 0 \quad \text{for any} \quad \bar{\u} \in \mathcal{D} \; \text{and} \; \bar{\y} \in [0, 1]^{n} 
\] 
where $\mathcal{D}$ is the dataset, and therefore, there will be no updates to the neural network weights $\tilde{\o}$ from the potential $\phi_j$. This loss of weight updates might lead to locally optimal solutions to the training optimization \eqref{eq:opt}, and that may be avoided by adding a penalty term to the optimization \eqref{eq:convex_difference} for each potential $\phi_j$ as 
\begin{align}\label{eq:regularization}
\psi_j (\u,\o, \y) = \mu \left( l_j(\p(\u;\o), \y)\right)^{2}\quad \text{with} \quad \mu > 0 
\end{align}
The penalty term $\psi_j$ penalizes the objective function, in case, the hinge loss potential $\phi_j$ does not trigger. The optimization \eqref{eq:convex_difference} with the regularizer is given by
\begin{align}
\w_t = \argmin_{\w}\; & \tilde{f}(\u, \w, \y_t - \alpha \nabla_{\y}L(\hat{\y},\y_t)) \nonumber \\ 
&- \tilde{f}(\u,\w, \y_t) + \mu \Omega (\u,\w, \y_t)\label{eq:convex_diff}
\end{align}
where 
\[
\Omega (\u,\w, \y_t) = \sum_{j=1}^{k} \psi_j (\u,\o, \y_t - \alpha \nabla_{\y}L(\hat{\y},\y_t)) - \psi_j (\u,\o, \y_t)  
\]
for $k$ potentials, and $\mu > 0$ . 

	These two optimizations \eqref{eq:inf} and \eqref{eq:convex_diff} are executed alternatively until convergence in Algorithm \ref{alg:opt}.
\begin{algorithm}
        \caption{Joint optimization: backpropagating loss to the neural network}\label{alg:opt}
        \begin{algorithmic}
        \STATE Initialization: $t = 1; \alpha, \eta, \mu > 0;  N, T \geq 1.$
        \sp
        \STATE Neural network weights $\w_0$ are initialized using standard techniques.
        \sp
                \WHILE{$ t \leq T$} 
                \sp
                    \STATE $ \y_{t} = \argmin_{\y} \,\, \tilde{f} (\u, \w_{t-1}, \y) $
                    \COMMENT{MAP inference}
                    \sp
                    \STATE $\w_t = \w_{t-1}$
                    \sp
                    \FOR {$i = 1,\dots,N$}
                        \STATE $\begin{aligned} \w_t \mathrel{-}=  \eta  \nabla_{\w_t} \big[& \tilde{f}(\u, \w_t, \y_t - \alpha \nabla_{\y}L(\hat{\y},\y_t)) \\
& - \tilde{f}(\u,\w_t, \y_t) + \mu \Omega(\u, \w_t,\y_t)\big]
\end{aligned}$
                    \ENDFOR
                \sp
                \STATE $t = t+1$
        \ENDWHILE
        \end{algorithmic}
\end{algorithm}

\subsection{Scalability}
DeepPSL fully inherits the scalability of PSL, that is, they have the same Big O behavior.The first step in DeepPSL inference consists of a forward pass through the NNs to compute the groundings, only adding time linear in number of ground terms. The second step is standard PSL inference, that is extremely efficient and scales linearly in number of potentials, see Section \ref{ssec:hlmrf}.

DeepPSL training requires alternate execution of gradient descent step for minimizing loss \eqref{eq:convex_diff} and an inference step. The gradient descent step can become memory intensive. As the loss function can be rewritten as summation of losses for each of the grounding, we address the problem using gradient accumulation. Weight updates are made after accumulating gradients from all the grounded potentials. Preserving the inherent scalability of PSL inference and addressing the challenges during training makes DeepPSL highly scalable.

\section{Experimental Evaluation}
We evaluate DeepPSL on three tasks -- T1: a digit addition task that shows that DeepPSL can learn predicates that are hard to specify in PSL, T2: a document classification relational problem that is of moderate scale and T3: a challenging large scale entity resolution problem. We compare our method with state-of-the-art neuro-symbolic methods and in some cases, with other methods that are more specific to the task. Each of these tasks contains train, validation and test splits in the corresponding dataset(s). We select the model and hyper-parameters, see Table \ref{tab:hyper-param}, that give the highest performance on the validation set. The selected model is then evaluated on the test set to report the performance. The performance metric is reported with a $95\%$ confidence interval, calculated by repeating each experiment multiple times. The experiments are performed on a MacBook Pro with 2.6GHz Intel i7 processor having 6 cores.
\begin{table}[H]
    \caption{Hyperparameters used for DeepPSL}
    \label{tab:hyper-param}
    \centering
\begin{threeparttable}
    \begin{tabular}{llll}
        \toprule
            Inference parameters  & \textbf{T1}  & \textbf{T2}  & \textbf{T3} \\
        \midrule
            Optimizer       & SGD  & SGD & SGD\\
            Learning rate    & $5\e{-3}$ & $5\e{-3}$ & $1\e{-2}$\\
            Loss change threshold & $1\e{-6}$ & $5\e{-3}$ & $1\e{-1}$ \\
            Max iterations  & 5000 & 1000 & 5000 \\
        $\underline{\gamma_i}$, $\overline{\gamma_i}$ & 20 & 20 & 20\\
        \midrule
           Training Parameters \tnote{$\star$}& \\
        \midrule
            Optimizer       & Adagrad & Adam & Adagrad\\
            $\alpha$   &  $5\e{-5}$ &  $1$ &  $5\e{-3}$\\
            $\mu$       & $1\e{-3}$ &  $12\e{-2}$ &  $0$\\
            Update steps ($N$) & 2 & 10 & 2\\
            Epochs ($T$)  & 10 & 75 & 100 \\
        \bottomrule
\end{tabular}
    \begin{tablenotes}
    \item[$\star$] {\footnotesize Rule weights are initialized randomly by drawing samples from a normal distribution $\mathcal{N}(1.0, 0.1)$.}
    \end{tablenotes}
\end{threeparttable}
\end{table}

\subsection{T1: Addition of handwritten digits} \label{exp_addition}
The goal of this task is to predict the sum of digits present in two MNIST images\footnote{A detailed explanation on digit addition problem may be found in \cite{DeepProbLog}.}  \cite{DeepProbLog}, a relatively simpler problem that is already well addressed by several neuro-symbolic methods. Note that the training data provides only the sum of the digits, and the digit labels are not provided. Hence, it is not possible to directly learn or specify a predicate to classify individual images, making it difficult for PSL to solve this problem. We investigate whether DeepPSL can learn the image predicate and achieve performance comparable to other neuro-symbolic approaches.

The datasets for this problem are generated following the procedure described in \cite{winters2021deepstochlog}. Rules provided in Figure \ref{rules:da} are used to add two digits in DeepPSL system. We use the predicate $\pslpred{Digit}$, a convolutional neural network (CNN), to recognize the digits present in the input images ($\pslarg{Img1}$, $\pslarg{Img2}$). DeepPSL is trained end-to-end to learn the rule weights as well as the CNN parameters, by minimizing the cross-entropy loss between the inferred sums ($\pslarg{S}$) and the ground truth.
 
The CNN consists of two convolution (CONV) layers with 32 and 64 filters of size 3 and stride 1 with ELU activation. Max pool layer with size 2 is applied on the output of last CONV layer. This is followed by two fully connected layers with 128 and 10 nodes on which ELU and softmax activations are applied respectively. A batch size of 16 is used for training. The learning rate for rule weights and CNN are decayed for the 10 epochs according to $\eta = \eta_0/(E+1)$ where, $\eta_0$ ($5\e{-4}$ for rule weights and  $1\e{-3}$ for CNN parameters) is the initial learning rate and $E$ is the epoch number (starting with 0).
A weight decay of $1\e{-7}$ is used for CNN parameters from the second epoch of training. Weight decay is not used for rule weights. In the ruleset, the summation constraint is implemented as a soft constraint with a fixed weight 10.
 
\begin{figure}[htpb]
    \centering
    \noindent\fbox{%
        \begin{minipage}{0.95\hsize}
            \begin{scriptsize}
        \vspace{-0.25cm}
            \begin{flalign*}
                \hspace{0.1cm} & \textit{\textbf{\# Addition of handwritten digits rules}}&& \\
                \theta_{1}:& \pslpred{Digit}(\pslarg{Img1}, \pslarg{D1}) \psland \pslpred{Digit}(\pslarg{Img2}, \pslarg{D2}) \psland (\pslarg{Img1}\pslneg = \pslarg{Img2}) \pslthen \pslpred{Sum} (\pslarg{S}) \; \widehat{} \; 2 \\
                \theta_{2}:& \pslpred{Digit}(\pslarg{Img1}, \pslarg{D1}) \psland \pslpred{Digit}(\pslarg{Img2}, \pslarg{D2}) \psland (\pslarg{Img1}\pslneg = \pslarg{Img2}) \pslthen \pslneg \pslpred{Sum} (\pslarg{S}) \; \widehat{} \; 2 \\[0.2cm]
                \hspace{0.2cm} & \textit{\textbf{\# Summation Constraint}}&& \\
                & \pslpred{Sum}(+\pslarg{S}) = 1  \\
            \end{flalign*}
            \end{scriptsize}
    \vspace{-0.9cm}
        \end{minipage}
    \hspace{-0.4cm}}
    \caption{DeepPSL rule set for addition of handwritten digits}
    \label{rules:da}
\end{figure}
The average classification accuracy of DeepPSL over 10 runs is shown in Table \ref{tab:addition-comparison}. The performance numbers for other methods are obtained from \cite{winters2021deepstochlog}. We also evaluate the CNN corresponding to \pslpred{Digit} predicate on the test split of MNIST data set. The CNN achieves an accuracy of $98.2\%$ demonstrating that DeepPSL could successfully learn the predicate even without any explicit data and thus achieves performance comparable to other neuro-symbolic methods.
 
\begin{table}[htb]
    \caption{Test accuracy on addition of handwritten digits}
    \label{tab:addition-comparison}
    \centering
    \begin{tabular}{l l l l @{\hspace{-0.12ex}}}
    \toprule
        NeurASP  & DeepProbLog & DeepStochLog & DeepPSL\\
        \midrule
        $97.3 \pm 0.3$ & $97.2 \pm 0.5$ & $97.9 \pm 0.1$ & $96.2 \pm 0.2$      \\
    \bottomrule
    \end{tabular}
\end{table}

\subsection{T2: Semi-supervised Classification}
The goal is to classify unlabeled documents in citation networks given some documents that are labeled. The problem is more challenging than task T1 as there is significantly more relational information available, which in turn poses a challenge to the scalability of neuro-symbolic methods. We use data from the Cora and Citeseer scientific datasets \cite{yang2016}. The Cora dataset contains 2708 documents in 7 categories, with 5429 citation links, and each document is represented by indicating the absence or presence of the corresponding word from a dictionary of 1433 unique words. Similarly, the Citeseer dataset contains 3327 documents in 6 categories, with 4591 citation links, and each document is represented by indicating the absence or presence of the corresponding word from a dictionary of 3703 unique words.

\begin{table}[htb]
    \caption{Dataset splits for semi-supervised classification task}
    \label{tab:nc_ds}
    \centering
    \begin{tabular}{llcc}
    \toprule
    Dataset       & Nodes & Split 1  &  Split 2  \\
                  &       & (Train/ Val/ Test) & (Train/ Val/ Test) \\
    \midrule
     Cora & 2708 &140/ 500/ 1000 & 1708/ 500/ 500     \\
     Citeseer & 3327 &120/ 500/ 1000 & 2327/ 500/ 500     \\
    \bottomrule
    \end{tabular}
\end{table}

	The predicate $\pslpred{Cite}(\pslarg{A},\pslarg{B})$ defines a citation from node $\pslarg{A}$ to node $\pslarg{B}$, and the number of neighbors of $\pslarg{A}$ are $n_{\pslarg{A}} = \left|\left\{\pslarg{B} \; \left| \; \pslpred{Cite} (\pslarg{B},\pslarg{A}) \right. \right\}\right|$. Furthermore, we define 2-hop neighbors and 3-hop neighbors as, for any $\pslarg{A} != \pslarg{B}$, 
\begin{align*}
\pslpred{CiteP}(\pslarg{A}, \pslarg{B}) &= \pslpred{Cite}(\pslarg{A},\pslarg{C}) \psland \pslpred{Cite}(\pslarg{C},\pslarg{B})\; \text{for any node } \pslarg{C}, \\
\pslpred{CiteQ}(\pslarg{A}, \pslarg{B}) &= \pslpred{CiteP}(\pslarg{A},\pslarg{C}) \psland \pslpred{Cite}(\pslarg{C},\pslarg{B}) \; \text{for any node } \pslarg{C}.
\end{align*}

In the rule set shown in Figure \ref{rules:nc}, the predicates \pslpred{Cite}, \pslpred{CiteP} and \pslpred{CiteQ} are directly observed from the data. The inferred predicate \pslpred{Label} identifies the labels of the documents. The deep learning predicates \pslpred{Neural} and \pslpred{Similar} represent the neural net classifier output and the similarity between documents, respectively.

\begin{figure}[htpb] 
    \centering
    \noindent\fbox{%
        \begin{minipage}{0.99\hsize}
            \begin{scriptsize}
	    \vspace{-0.25cm}
            \begin{flalign*}
                \hspace{0.2cm} & \textit{\textbf{\# Semi-supervised classification rules}}&& \\
                \theta_{1}:& \pslpred{Neural}(\pslarg{A}, \pslarg{Y}) \pslthen \pslpred{Label}(\pslarg{A}, \pslarg{Y}) \\
                \frac{\theta_{2}}{n_{\pslarg{A}}}:& \pslpred{Label}(\pslarg{B}, \pslarg{Y}) \psland \pslpred{Similar}(\pslarg{B}, \pslarg{A}) \psland \pslpred{Cite}(\pslarg{B}, \pslarg{A})\pslthen \pslpred{Label}(\pslarg{A}, \pslarg{Y}) \\
                \frac{\theta_{3}}{n_{\pslarg{A}}}:& \pslpred{Label}(\pslarg{B}, \pslarg{Y}) \psland \pslneg \pslpred{Similar}(\pslarg{B}, \pslarg{A}) \psland \pslpred{Cite}(\pslarg{B}, \pslarg{A})\pslthen \pslneg \pslpred{Label}(\pslarg{A}, \pslarg{Y}) \\
                \theta_{4}:& \frac{1}{|\pslarg{B}|}\pslpred{Neural}(+\pslarg{B}, \pslarg{Y}) \pslthen \pslpred{Label}(\pslarg{A}, \pslarg{Y}) \{ \pslarg{B}: \pslpred{Cite}(\pslarg{A}, \pslarg{B}) \} \\
                \theta_{5}:& \frac{1}{|\pslarg{B}|}\pslpred{Neural}(+\pslarg{B}, \pslarg{Y}) \pslthen \pslpred{Label}(\pslarg{A}, \pslarg{Y}) \{ \pslarg{B}: \pslpred{CiteP}(\pslarg{A}, \pslarg{B}) \} \\
                \theta_{6}:& \frac{1}{|\pslarg{B}|}\pslpred{Neural}(+\pslarg{B}, \pslarg{Y}) \pslthen \pslpred{Label}(\pslarg{A}, \pslarg{Y}) \{ \pslarg{B}: \pslpred{CiteQ}(\pslarg{A}, \pslarg{B}) \} \\[0.2cm]
               & \textit{\textbf{\# Constraints}} \\
               \phantom{w_{4}:}& \pslpred{Label}(\pslarg{A}, +\pslarg{Y}) =1\\
            \end{flalign*}
	    \vspace{-0.9cm}
            \end{scriptsize}
        \end{minipage}
    \hspace{-0.4cm}}
    \caption{DeepPSL rule set for semi-supervised classification}
    \label{rules:nc}
\end{figure}
The predicate \pslpred{Neural} is represented by a feedforward neural network with an input layer that takes the document features, followed by a hidden layer with 16 nodes (RELU activation), a drop out layer with a rate of 0.2, and a final softmax layer with 7 nodes corresponding to the output classes. The predicate \pslpred{Label} is represented by a Siamese network composed of two identical networks that share the first layer of the feedforward network, and a distance layer with 16 nodes. We minimize cross entropy loss and use learning rate $2\e{-3}$ for rule weights and $2\e{-2}$ for NN parameters. The weight decay parameter of Adam optimizer is set to $3\e{-4}$ (Split 1) and $3\e{-5}$ (Split 2), and is only used for NN parameters. The summation constraint is implemented as a soft constraint with a fixed weight 20.    

We compare the performance of DeepPSL against various baselines on the Cora and Citeseer datasets, using randomly generated data splits described in Table \ref{tab:nc_ds}. The results, presented in Table \ref{tab:node_acc}, show classification class averaged accuracy on the test sets. The results for GCN\footnote{https://github.com/tkipf/gcn} \cite{gcn}, PSL and DeepPSL are generated by running each method 100 times, while the results for ManiReg \cite{mani_reg}, SemiEmb \cite{semi_emb}, LP \cite{lp}, DeepWalk \cite{deep_walk}, ICA \cite{ica}, Planetoid \cite{yang2016} are taken from \cite{gcn}. As DeepStochLog training was slow, we ran it only 5 times and don’t report error bars. Both splits are evaluated for GCN, PSL and DeepPSL while other results are reported only for Split 1.

DeepPSL achieves the highest accuracy for both data sets and both splits, outperforming DeepStochLog by a large margin and surpassing even methods that are specialized for semi-supervised learning. The expressivity of PSL for relational systems enables DeepPSL to leverage the labels and features of its neighbors, while the end-to-end training helps optimize the weights of the neural predicates to maximize classification accuracy. When compared to other neuro-symbolic methods, DeepPSL demonstrates excellent scalability. DeepProbLog , timed out for both networks while DeepStochLog training time was $\sim 50$x that of DeepPSL.

\begin{table}[htb]
    \caption{Classification accuracy on test nodes for Task T2}
    \label{tab:node_acc}
    \centering
    \begin{tabular}{lllll}
    \toprule
    Algorithm       & Cora & Citeseer \\
    \midrule
    ManiReg  & $59.5$ & $60.1$ \\
    SemiEmb  & $59.0$ & $59.6$ \\
    LP & $68.0$ & $45.3$  \\
    DeepWalk  & $67.2$ & $43.2$ \\
    ICA & $75.1$ & $69.1$ \\  
    Planetoid & $75.7$ & $64.7$ \\
    DeepStochLog    & $69.4$ & $65$ \\
    DeepProbLog    & timeout & timeout \\
    GCN    & $80.08 \pm 0.34$  & $67.96 \pm 0.32$ \\
    PSL   & $62.97 \pm 0.52$  & $64.88 \pm 0.38$ \\
    DeepPSL & \bm{$81.31 \pm 0.28$}  & \bm{$69.11 \pm 0.27$} \\
    \midrule
    GCN(Split 2)    & $87.46 \pm 0.30$  & $75.96 \pm 0.34$ \\
    PSL(Split 2)   & $85.94 \pm 0.28$  & $75.66 \pm 0.32$ \\
    DeepPSL(Split 2)& \bm{$88.94 \pm 0.25$}  & \bm{$76.01 \pm 0.31$} \\
    \bottomrule
    \end{tabular}
\end{table}

\subsection{T3: Entity Resolution} \label{exp_ER}
We perform entity resolution on CiteSeer database \cite{bhattacharya:tkdd07} using DeepPSL to identify duplicate references to authors and published papers.  As the task requires predicting the edges between every pair of author nodes and paper nodes in a large network, the problem results in millions of ground rules posing a major challenge for existing neuro-symbolic approaches.

The data consists of author names, paper titles and relational information such as authorship of papers. There are around 3000 author references and 1500 paper references. We use the train and test splits provided in PSL entity resolution example.\footnote{\label{psl_repo}\url{https://github.com/linqs/psl-examples/tree/master/entity-resolution}} We extract a third of data in the provided train set to create a validation set.
 
For this problem, pair-wise similarity of author names and pair-wise similarity of paper titles provide key information to identify duplicates. Traditionally, this similarity is computed with hand crafted metrics and is used for inference. It has been shown that the performance of different standard string similarity metrics varies greatly based on the application domain \cite{StrSimOntAlign}. With DeepPSL, we do not rely on pre-determined string similarity metrics. Rather, the system learns the optimal way to compute similarities from the provided application specific data.

\begin{figure}[htpb]
    \centering
    \noindent\fbox{%
        \begin{minipage}{0.95\hsize}
            \begin{scriptsize}
        \vspace{-0.25cm}
            \begin{flalign*}
               \hspace{0.2cm} & \textit{\textbf{\# Entity resolution rules}}&& \\
                \theta_{1}:& \pslpred{AuthorName}(\pslarg{A1}, \pslarg{N1}) \psland \pslpred{AuthorName}(\pslarg{A2}, \pslarg{N2}) \psland \pslpred{SimName}(\pslarg{N1}, \pslarg{N2}) \\ & \pslthen \pslpred{SameAuthor}(\pslarg{A1}, \pslarg{A2})\; \widehat{} \; 2 \\
                \theta_{2}:& \pslpred{PaperTitle}(\pslarg{P1}, \pslarg{T1}) \psland \pslpred{PaperTitle}(\pslarg{P2}, \pslarg{T2})\psland \pslpred{SimTitle}(\pslarg{T1}, \pslarg{T2})\\ &\pslthen \pslpred{SamePaper}(\pslarg{P1}, \pslarg{P2})\; \widehat{} \; 2 \\
                \theta_{3}:& \pslpred{AuthorBlk}(\pslarg{A1}, \pslarg{B1}) \psland \pslpred{AuthorBlk}(\pslarg{A2}, \pslarg{B1})  \psland \pslpred{AuthorOf}(\pslarg{A1}, \pslarg{P1})\psland \\ &\pslpred{AuthorOf}(\pslarg{A2}, \pslarg{P2}) \psland \pslpred{SamePaper}(\pslarg{P1}, \pslarg{P2})\pslthen \pslpred{SameAuthor}(\pslarg{A1}, \pslarg{A2}) \; \widehat{} \; 2\\
                \theta_{4}:& \pslpred{AuthorBlk}(\pslarg{A1}, \pslarg{B}) \psland \pslpred{AuthorBlk}(\pslarg{A2}, \pslarg{B}) \psland \pslpred{AuthorBlk}(\pslarg{A3}, \pslarg{B}) \psland \\
                                       &\pslpred{SameAuthor}(\pslarg{A1}, \pslarg{A2}) \psland \pslpred{SameAuthor}(\pslarg{A2}, \pslarg{A3}) \psland (\pslarg{A1} \pslneg = \pslarg{A3})\psland\\
                                       &(\pslarg{A2} \pslneg = \pslarg{A3}) \psland (\pslarg{A1} \pslneg = \pslarg{A2}) \pslthen \pslpred{SameAuthor}(\pslarg{A1}, \pslarg{A3}) \; \widehat{} \; 2\\
                \theta_{5}:& \pslneg \pslpred{SameAuthor}(\pslarg{A1}, \pslarg{A2}) \; \widehat{} \; 2 \\
                \theta_{6}:& \pslneg \pslpred{SamePaper}(\pslarg{P1}, \pslarg{P2}) \; \widehat{} \; 2 \\[0.2cm]
                \hspace{0.2cm} & \textit{\textbf{\# Identity Constraints}}&& \\
                & \pslpred{SameAuthor}(\pslarg{A}, \pslarg{A}) = 1  \\
                & \pslpred{SamePaper}(\pslarg{P}, \pslarg{P}) = 1  \\
            \end{flalign*}
        \vspace{-0.9cm}
            \end{scriptsize}
        \end{minipage}
    \hspace{-0.4cm}}
    \caption{DeepPSL rule set for entity resolution}
    \label{rules:er}
\end{figure}
Rules\textsuperscript{\ref{psl_repo}} in Figure \ref{rules:er} incorporate the author name and paper title similarity values in conjunction with other relational information to identify duplicates. The inferred predicates $\pslpred{SameAuthor}$ and $\pslpred{SamePaper}$ identify if authors and papers are same respectively. $\pslpred{AuthorName}$, $\pslpred{PaperTitle}$, $\pslpred{AuthorOf}$ and $\pslpred{AuthorBlock}$ are directly observed from the data. In the DeepPSL system, we associate author name similarity predicate $\pslpred{SimName}$ and title similarity predicate $\pslpred{SimTitle}$ with siamese networks\cite{siamese}.
The NN to compute author name similarity takes 56 dimensional character ([A-Za-z .,']) based one hot encoding of the names as input. NN for title similarity operates on mean of 300 dimensional vectors derived from GloVe embeddings (pre-trained on Wikipedia 2014 and Gigaword5 corpus.\footnote{\url{https://nlp.stanford.edu/projects/glove}}) of the words in the title. Each twin in the architecture has a hidden layer of 50 nodes and distance layer of 50 nodes for both NNs. We minimize cross entropy loss to learn rule weights and weights of NNs for author name and paper title similarities. During training of the DeepPSL system, we do not perform any sampling over the graph. We use learning rate of $5\e{-2}$ for NN parameters and $25\e{-3}$ for rule weights. Weight decay is not used for rule weights but for NN parameters, $1\e{-3}$ is used. The identity constraints are implemented with a fixed weight 10. The model which gives the highest average of F1-scores for same author and same paper identification is selected based on validation set.

After trivial potential removal, there are 6.5 million and 3.1 million ground rules during train and test respectively. The scale of this relational problem is out of scope for neural PLP approaches such as DeepProbLog. We attempted a comparison with DeepStochLog\footnote{ \url{https://github.com/ML KULeuven/deepstochlog/tree/main/examples}.}. However, DeepStochLog could not scale to the size of this problem, encountering memory errors and timing out even when using only a subset of the rules.
We report performance of DeepPSL and PSL over 10 runs in Table \ref{er-performance-table}. PSL setup is built with same ruleset as in DeepPSL but the author name and paper title similarities are computed using metrics mentioned in \cite{bhattacharya:tkdd07}. PSL-A uses Soft TFIDF with Jaro-Winkler \cite{SoftTFIDF} for both author name similarity and paper title similarity. PSL-B uses Soft TFIDF for author names and cosine similarity of GloVe based vectors for paper titles.
We also experiment with PSL setups using other standard similarity metrics (as mentioned in \cite{StrSimOntAlign,SimMTitle,ImpSimMShortText}): Cosine, Jaccard, Monge Elkan (with Levenshtein). The PSL systems corresponding to these metrics are designated as PSL-C, PSL-J and PSL-M respectively. In each of these setups, we use the same similarity metric for computing author name and paper title similarities. Strings are not preprocessed apart from tokenization when necessary. Error bars are not reported for PSL as there is insignificant variation across runs.
\begin{table}[htb]
    \caption{Performance on Task T3}
    \label{er-performance-table}
    \centering
    \begin{tabular}{lll}
    \toprule
    Algorithm & Author F1 score& Paper F1 score\\
    \midrule
    PSL-A  &  $0.9271$  & $0.8276$ \\
    PSL-B  &  $0.8958$  & $0.8501$ \\
    PSL-C  &  $0.7852$  & $0.7656$ \\
    PSL-J  &  $0.8073$  & $0.7631$ \\
    PSL-M  &  $0.8008$  & $0.8884$ \\
    DeepStochLog  &  timeout  & timeout \\
    \textbf{DeepPSL} & $\mathbf{0.9468 \pm 0.0061}$ & $\mathbf{0.9228 \pm 0.0036}$ \\
    \bottomrule
   \end{tabular}
\end{table}
DeepPSL outperforms all the PSL setups for both author and paper entity resolution, see Table \ref{er-performance-table}. The PSL performance depends on the pre-determined similarity metric used and it is hard to find a similarity metric that yields the highest accuracy. DeepPSL achieves excellent performance by directly learning optimal similarity functions for the entity resolution task. Moreover, these results highlight that DeepPSL scales to a difficult problem which proves to be out of reach for competing methods.

\section{Conclusions and Future Work}
We introduced DeepPSL an end-to-end trainable system that integrates reasoning and perception. We proposed a novel algorithm to enable end-to-end training. Experimental results on three different tasks demonstrated the broad applicability of the method. DeepPSL scales to relational problems that prove to be challenging for competing methods. DeepPSL has some limitations. Firstly, as DeepPSL uses deep networks to model predicates, learning is not convex, and therefore, the proposed approach may suffer from local minima. While we did not observe any significant sensitivity to local minima in our experiments, further research is needed to understand this better. Secondly, the work described in this article does not address latent variables. Future work will report on extensions of DeepPSL to the latent variable case.


\section*{Disclaimer} The views reflected in this article are the views of the authors and do not necessarily reflect the views of the global EY organization or its member firms.

\bibliographystyle{named}
\bibliography{references}

\begin{thebibliography}{}

\bibitem[\protect\citeauthoryear{Agrawal \bgroup \em et al.\egroup
  }{2019}]{AgarwalAmosBarratt}
Akshay Agrawal, Brandon Amos, Shane Barratt, Stephen Boyd, Steven Diamond, and
  J.~Zico Kolter.
\newblock Differentiable {C}onvex {O}ptimization {L}ayers.
\newblock In {\em Advances in Neural Information Processing Systems},
  volume~32. Curran Associates Inc., 2019.

\bibitem[\protect\citeauthoryear{Agrawal \bgroup \em et al.\egroup
  }{2020}]{AgarwalBarrattBoyd}
Akshay Agrawal, Shane Barratt, Stephen Boyd, Enzo Busseti, and Walaa~M. Moursi.
\newblock Differentiating {T}hrough a {C}one {P}rogram.
\newblock {\em arXiv; 1904.09043}, 2020.

\bibitem[\protect\citeauthoryear{Amos and Kolter}{2017}]{AmosKolter}
Brandon Amos and J~Zico Kolter.
\newblock Optnet: {D}ifferentiable {O}ptimization as a {L}ayer in {N}eural
  {N}etworks.
\newblock In {\em International Conference on Machine Learning}, pages
  136--145. PMLR, 2017.

\bibitem[\protect\citeauthoryear{Augustine and Getoor}{2018}]{trivialblocking}
E.~Augustine and L.~Getoor.
\newblock {A Comparison of Bottom-Up Approaches to Grounding for Templated
  Markov Random Fields}.
\newblock In {\em SysML}, 2018.

\bibitem[\protect\citeauthoryear{Bach \bgroup \em et al.\egroup
  }{2017}]{bach2017hinge}
Stephen~H Bach, Matthias Broecheler, Bert Huang, and Lise Getoor.
\newblock Hinge-{L}oss {M}arkov {R}andom {F}ields and {P}robabilistic {S}oft
  {L}ogic.
\newblock {\em Journal of Machine Learning Research}, 18:1--67, 2017.

\bibitem[\protect\citeauthoryear{Belkin \bgroup \em et al.\egroup
  }{2006}]{mani_reg}
Mikhail Belkin, Partha Niyogi, and Vikas Sindhwani.
\newblock Manifold regularization: {A} geometric framework for learning from
  labeled and unlabeled examples.
\newblock {\em Journal of machine learning research}, 7(11), 2006.

\bibitem[\protect\citeauthoryear{Bhattacharya and
  Getoor}{2007}]{bhattacharya:tkdd07}
Indrajit Bhattacharya and Lise Getoor.
\newblock Collective {E}ntity {R}esolution {I}n {R}elational {D}ata.
\newblock {\em ACM Transactions on Knowledge Discovery from Data}, pages 1--36,
  2007.

\bibitem[\protect\citeauthoryear{Cheatham and Hitzler}{2013}]{StrSimOntAlign}
Michelle Cheatham and Pascal Hitzler.
\newblock String {S}imilarity {M}etrics for {O}ntology {A}lignment.
\newblock In {\em The Semantic Web -- ISWC 2013}, pages 294--309, Berlin,
  Heidelberg, 2013. Springer Berlin Heidelberg.

\bibitem[\protect\citeauthoryear{Cohen \bgroup \em et al.\egroup
  }{2003}]{SoftTFIDF}
W.~Cohen, P.~Ravikumar, and S.~Fienberg.
\newblock A {C}omparison of {S}tring {D}istance {M}etrics for {N}ame-{M}atching
  {T}asks.
\newblock In {\em The IJCAI Workshop on Information Integration on the Web
  (IIWeb)}, 2003.

\bibitem[\protect\citeauthoryear{Cohen \bgroup \em et al.\egroup
  }{2020}]{TensorLog}
William~W. Cohen, Fan Yang, and Kathryn Mazaitis.
\newblock Tensor{L}og: {A} {P}robabilistic {D}atabase {I}mplemented {U}sing
  {D}eep-{L}earning {I}nfrastructure.
\newblock {\em Journal of Artificial Intelligence Research}, 67:285--325, 2020.

\bibitem[\protect\citeauthoryear{Dempe}{2018}]{dempe2018bilevel}
Stephan Dempe.
\newblock {\em Bilevel {O}ptimization: {T}heory, {A}lgorithms and
  {A}pplications}.
\newblock TU Bergakademie Freiberg, Fakult{\"a}t f{\"u}r Mathematik und
  Informatik, 2018.

\bibitem[\protect\citeauthoryear{Diligenti \bgroup \em et al.\egroup
  }{2017}]{semantic_regularization}
Michelangelo Diligenti, Marco Gori, and Claudio Sacc{\`{a}}.
\newblock Semantic-based regularization for learning and inference.
\newblock {\em Artificial Intelligence}, 244:143--165, 2017.

\bibitem[\protect\citeauthoryear{Donadello \bgroup \em et al.\egroup
  }{2017}]{Logic_Tensor_Networks}
Ivan Donadello, Luciano Serafini, and Artur~D'Avila Garcez.
\newblock Logic {T}ensor {N}etworks for {S}emantic {I}mage {I}nterpretation.
\newblock In {\em Proceedings of the 26th International Joint Conference on
  Artificial Intelligence}, IJCAI'17, page 1596–1602. AAAI Press, 2017.

\bibitem[\protect\citeauthoryear{Dong \bgroup \em et al.\egroup
  }{2019}]{Neural_Logic_Machines}
Honghua Dong, Jiayuan Mao, Tian Lin, Chong Wang, Lihong Li, and Denny Zhou.
\newblock Neural {L}ogic {M}achines.
\newblock In {\em International Conference on Learning Representations}, 2019.

\bibitem[\protect\citeauthoryear{Farnadi \bgroup \em et al.\egroup
  }{2018}]{fairness18}
Golnoosh Farnadi, Behrouz Babaki, and Lise Getoor.
\newblock {Fairness in Relational Domains}.
\newblock In {\em Proceedings of the 2018 AAAI/ACM Conference on AI, Ethics,
  and Society}, page 108–114, 2018.

\bibitem[\protect\citeauthoryear{Gali \bgroup \em et al.\egroup
  }{2016}]{SimMTitle}
Najlah Gali, Radu Mariescu-Istodor, and Pasi Franti.
\newblock Similarity {M}easures for {T}itle {M}atching.
\newblock In {\em International Conference on Pattern Recognition (ICPR)},
  2016.

\bibitem[\protect\citeauthoryear{Gandoura \bgroup \em et al.\egroup
  }{2020}]{gandoura2020human}
Marouene~Sfar Gandoura, Zografoula Vagena, and Nikolaos Vasiloglou.
\newblock {Human in the Loop Enrichment of Product Graphs with Probabilistic
  Soft Logic}.
\newblock In {\em Proceedings of Knowledge Graphs and E-commerce, KDD 20},
  2020.

\bibitem[\protect\citeauthoryear{Ghadimi and
  Wang}{2018}]{ghadimi2018approximation}
Saeed Ghadimi and Mengdi Wang.
\newblock Approximation {M}ethods for {B}ilevel {P}rogramming.
\newblock {\em arXiv preprint arXiv:1802.02246}, 2018.

\bibitem[\protect\citeauthoryear{Gould \bgroup \em et al.\egroup
  }{2016}]{GouldFernando}
Stephen Gould, Basura Fernando, Anoop Cherian, Peter Anderson, Rodrigo~Santa
  Cruz, and Edison Guo.
\newblock On {D}ifferentiating {P}arameterized {A}rgmin and {A}rgmax {P}roblems
  with {A}pplication to {B}i-level {O}ptimization.
\newblock {\em arXiv; 1607.05447}, 2016.

\bibitem[\protect\citeauthoryear{Gridach}{2020}]{PSL_SL_Neural}
Mourad Gridach.
\newblock A framework based on (probabilistic) soft logic and neural network
  for {NLP}.
\newblock {\em Applied Soft Computing}, 93:106232, 2020.

\bibitem[\protect\citeauthoryear{Hu \bgroup \em et al.\egroup
  }{2016}]{Logic_DL}
Zhiting Hu, Xuezhe Ma, Zhengzhong Liu, Eduard Hovy, and Eric Xing.
\newblock Harnessing {D}eep {N}eural {N}etworks with {L}ogic {R}ules.
\newblock In {\em Proceedings of the 54th Annual Meeting of the Association for
  Computational Linguistics (Volume 1: Long Papers)}, pages 2410--2420, Berlin,
  Germany, August 2016. Association for Computational Linguistics.

\bibitem[\protect\citeauthoryear{Jordan and Dimakis}{2020}]{jordan20}
Matt Jordan and Alexandros~G Dimakis.
\newblock Exactly {C}omputing the {L}ocal {L}ipschitz {C}onstant of {R}e{LU}
  {N}etworks.
\newblock In H.~Larochelle, M.~Ranzato, R.~Hadsell, M.~F. Balcan, and H.~Lin,
  editors, {\em Advances in Neural Information Processing Systems}, volume~33,
  pages 7344--7353, 2020.

\bibitem[\protect\citeauthoryear{Kipf and Welling}{2017}]{gcn}
Thomas~N. Kipf and Max Welling.
\newblock {S}emi-{S}upervised {C}lassification with {G}raph {C}onvolutional
  {N}etworks.
\newblock In {\em 5th International Conference on Learning Representations,
  {ICLR} 2017}. OpenReview.net, 2017.

\bibitem[\protect\citeauthoryear{Koch \bgroup \em et al.\egroup
  }{2015}]{siamese}
Gregory Koch, Richard Zemel, and Ruslan Salakhutdinov.
\newblock Siamese {N}eural {N}etworks for {O}ne-shot {I}mage {R}ecognition.
\newblock In {\em ICML 2015 Deep Learning Worshop}, 2015.

\bibitem[\protect\citeauthoryear{Koller \bgroup \em et al.\egroup
  }{2007}]{koller2007graphical}
Daphne Koller, Nir Friedman, Lise Getoor, and Ben Taskar.
\newblock {Graphical Models in a Nutshell}.
\newblock {\em Introduction to statistical relational learning}, 43, 2007.

\bibitem[\protect\citeauthoryear{Kouki \bgroup \em et al.\egroup
  }{2017}]{kouki2017collective}
Pigi Kouki, Jay Pujara, Christopher Marcum, Laura Koehly, and Lise Getoor.
\newblock Collective entity resolution in familial networks.
\newblock In {\em 2017 IEEE International Conference on Data Mining (ICDM)},
  pages 227--236. IEEE, 2017.

\bibitem[\protect\citeauthoryear{Lee \bgroup \em et al.\egroup
  }{2019}]{LeeMajiRavichandran}
Kwonjoon Lee, Subhransu Maji, Avinash Ravichandran, and Stefano Soatto.
\newblock Meta-{L}earning with {D}ifferentiable {C}onvex {O}ptimization.
\newblock In {\em Proceedings of the IEEE/CVF Conference on Computer Vision and
  Pattern Recognition}, pages 10657--10665, 2019.

\bibitem[\protect\citeauthoryear{Lu and Getoor}{2003}]{ica}
Qing Lu and Lise Getoor.
\newblock Link-based classification.
\newblock In {\em Proceedings of the 20th International Conference on Machine
  Learning (ICML-03)}, pages 496--503. AAAI Press, 2003.

\bibitem[\protect\citeauthoryear{Manhaeve \bgroup \em et al.\egroup
  }{2018}]{DeepProbLog}
Robin Manhaeve, Sebastijan Dumancic, Angelika Kimmig, Thomas Demeester, and Luc
  De~Raedt.
\newblock Deep{P}rob{L}og: {N}eural {P}robabilistic {L}ogic {P}rogramming.
\newblock In {\em Advances in Neural Information Processing Systems},
  volume~31. Curran Associates, Inc., 2018.

\bibitem[\protect\citeauthoryear{Marra \bgroup \em et al.\egroup
  }{2019}]{Deep_Logic_Models}
Giuseppe Marra, Francesco Giannini, Michelangelo Diligenti, and Marco Gori.
\newblock Integrating {L}earning and {R}easoning with {D}eep {L}ogic {M}odels.
\newblock In {\em Machine Learning and Knowledge Discovery in Databases -
  European Conference, 2019, Proceedings, Part {II}}, volume 11907, pages
  517--532. Springer, 2019.

\bibitem[\protect\citeauthoryear{Marra \bgroup \em et al.\egroup
  }{2020}]{marrarelational}
Giuseppe Marra, Michelangelo Diligenti, Francesco Giannini, Marco Gori, and
  Marco Maggini.
\newblock Relational {N}eural {M}achines.
\newblock In {\em 24th European Conference on Artificial Intelligence}, 2020.

\bibitem[\protect\citeauthoryear{Perera \bgroup \em et al.\egroup
  }{2018}]{perera2018cyberattack}
Ian Perera, Jena Hwang, Kevin Bayas, Bonnie Dorr, and Yorick Wilks.
\newblock {Cyberattack Prediction Through Public Text Analysis and
  Mini-Theories}.
\newblock In {\em 2018 IEEE International Conference on Big Data (Big Data)},
  pages 3001--3010. IEEE, 2018.

\bibitem[\protect\citeauthoryear{Perozzi \bgroup \em et al.\egroup
  }{2014}]{deep_walk}
Bryan Perozzi, Rami Al-Rfou, and Steven Skiena.
\newblock Deepwalk: {O}nline learning of social representations.
\newblock In {\em Proceedings of the 20th ACM SIGKDD international conference
  on Knowledge discovery and data mining}, pages 701--710, 2014.

\bibitem[\protect\citeauthoryear{Pryor \bgroup \em et al.\egroup
  }{2022}]{neu_psl}
Connor Pryor, Charles Dickens, Eriq Augustine, Alon Albalak, William Wang, and
  Lise Getoor.
\newblock {NeuPSL}: {N}eural {P}robabilistic {S}oft {L}ogic.
\newblock {\em arXiv preprint arXiv:2205.14268}, 2022.

\bibitem[\protect\citeauthoryear{Raedt \bgroup \em et al.\egroup
  }{2007}]{ProbLog}
Luc~De Raedt, Angelika Kimmig, and Hannu Toivonen.
\newblock Prob{L}og: {A} {P}robabilistic {P}rolog and {I}ts {A}pplication in
  {L}ink {D}iscovery.
\newblock In {\em 30th International Joint Conference on Artificial
  Intelligence}, pages 2462--2467, 2007.

\bibitem[\protect\citeauthoryear{Richardson and Domingos}{2006}]{MLN}
Matthew Richardson and Pedro Domingos.
\newblock Markov {L}ogic {N}etworks.
\newblock {\em Machine Learning}, 62(1–2):107–136, February 2006.

\bibitem[\protect\citeauthoryear{Rockafellar}{1981}]{roc81}
RT~Rockafellar.
\newblock Favorable {C}lasses of {L}ipschitz {C}ontinuous {F}unctions in
  {S}ubgradient {O}ptimization.
\newblock 1981.

\bibitem[\protect\citeauthoryear{Rockt{\"a}schel and
  Riedel}{2016}]{neural_theorem_provers}
Tim Rockt{\"a}schel and Sebastian Riedel.
\newblock Learning {K}nowledge {B}ase {I}nference with {N}eural {T}heorem
  {P}rovers.
\newblock In {\em Proceedings of the 5th Workshop on Automated Knowledge Base
  Construction}, pages 45--50, San Diego, CA, June 2016. Association for
  Computational Linguistics.

\bibitem[\protect\citeauthoryear{Rodden \bgroup \em et al.\egroup
  }{2020}]{rodden2020vmi}
Aaron Rodden, Tarun Salh, Eriq Augustine, and Lise Getoor.
\newblock {VMI-PSL: Visual Model Inspector for Probabilistic Soft Logic}.
\newblock In {\em Fourteenth ACM Conference on Recommender Systems}, pages
  604--606, 2020.

\bibitem[\protect\citeauthoryear{Rospocher}{2018}]{rospocher2018ontology}
Marco Rospocher.
\newblock {An Ontology-Driven Probabilistic Soft Logic Approach to Improve NLP
  Entity Annotations}.
\newblock In {\em International Semantic Web Conference}, pages 144--161.
  Springer, 2018.

\bibitem[\protect\citeauthoryear{Scaman and
  Virmaux}{2018}]{scaman2018lipschitz}
Kevin Scaman and Aladin Virmaux.
\newblock Lipschitz regularity of deep neural networks: analysis and efficient
  estimation.
\newblock In {\em Proceedings of the 32nd International Conference on Neural
  Information Processing Systems}, pages 3839--3848, 2018.

\bibitem[\protect\citeauthoryear{Sinha \bgroup \em et al.\egroup
  }{2017}]{sinha2017review}
Ankur Sinha, Pekka Malo, and Kalyanmoy Deb.
\newblock A {R}eview on {B}ilevel {O}ptimization: {F}rom {C}lassical to
  {E}volutionary {A}pproaches and {A}pplications.
\newblock {\em IEEE Transactions on Evolutionary Computation}, 22(2):276--295,
  2017.

\bibitem[\protect\citeauthoryear{Srinivasan \bgroup \em et al.\egroup
  }{2020}]{srini20}
Sriram Srinivasan, Eriq Augustine, and Lise Getoor.
\newblock {Tandem Inference: An Out-of-Core Streaming Algorithm for Very
  Large-Scale Relational Inference}.
\newblock {\em Proceedings of the AAAI Conference on Artificial Intelligence},
  34(06):10259--10266, 2020.

\bibitem[\protect\citeauthoryear{Weston \bgroup \em et al.\egroup
  }{2012}]{semi_emb}
Jason Weston, Fr{\'e}d{\'e}ric Ratle, Hossein Mobahi, and Ronan Collobert.
\newblock Deep learning via semi-supervised embedding.
\newblock {\em In Neural Networks: Tricks of the Trade}, pages 639--655, 2012.

\bibitem[\protect\citeauthoryear{Winters \bgroup \em et al.\egroup
  }{2022}]{winters2021deepstochlog}
Thomas Winters, Giuseppe Marra, Robin Manhaeve, and Luc~De Raedt.
\newblock Deepstochlog: Neural stochastic logic programming.
\newblock {\em Proceedings of the AAAI Conference on Artificial Intelligence},
  36(9):10090--10100, Jun. 2022.

\bibitem[\protect\citeauthoryear{Yang \bgroup \em et al.\egroup
  }{2016}]{yang2016}
Zhilin Yang, William Cohen, and Ruslan Salakhudinov.
\newblock Revisiting semi-supervised learning with graph embeddings.
\newblock In {\em International conference on machine learning}, pages 40--48.
  PMLR, 2016.

\bibitem[\protect\citeauthoryear{Yang \bgroup \em et al.\egroup
  }{2020}]{neur_asp}
Zhun Yang, Adam Ishay, and Joohyung Lee.
\newblock {NeurASP}: {E}mbracing {N}eural networks into {A}nswer {S}et
  {P}rogramming.
\newblock In {\em Proceedings of the Twenty-Ninth International Joint
  Conference on Artificial Intelligence, {IJCAI-20}}, pages 1755--1762.
  International Joint Conferences on Artificial Intelligence Organization, 7
  2020.

\bibitem[\protect\citeauthoryear{Yih and Meek}{2007}]{ImpSimMShortText}
Wen-Tau Yih and Christopher Meek.
\newblock Improving {S}imilarity {M}easures for {S}hort {S}egments of {T}ext.
\newblock In {\em Proceedings of the 22nd National Conference on Artificial
  Intelligence - Volume 2}, AAAI'07, page 1489–1494. AAAI Press, 2007.

\bibitem[\protect\citeauthoryear{Zhu \bgroup \em et al.\egroup }{2003}]{lp}
Xiaojin Zhu, Zoubin Ghahramani, and John~D Lafferty.
\newblock Semi-supervised learning using gaussian fields and harmonic
  functions.
\newblock In {\em Proceedings of the 20th International conference on Machine
  learning (ICML-03)}, pages 912--919, 2003.

\end{thebibliography}

\maketitle
\appendix
\section{DeepPSL Scalability} 
\begin{figure}[hbt]
\centering
\begin{tikzpicture}[scale=0.8]
\begin{axis}[
	legend style={at={(0.03,0.97)}, anchor=north west},
	xlabel={No of ground rules (in thousands)},    
	ylabel={Runtime per epochs (in $\sec$)},    
	xtick={30, 60, 120, 190, 280},
	ytick={0, 2, 4, 6, 8, 10, 12},
	    ]
\addplot+[red, mark options={red}, line width=0.8pt] coordinates {
    (30,2.08)
    (60,3.43)
    (120,6.42)
    (190,9.85)
    (280,11.75)
};
\addlegendentry{{\tiny Total training time}}

\addplot+[blue, mark options={blue}, line width=0.8pt] coordinates {
    (30,1.98)
    (60,3.26)
    (120,6.06)
    (190,9.15)
    (280,10.73)
};
\addlegendentry{{\tiny DeepPSL Inference time}}

\addplot+[cyan, mark options={cyan}, line width=0.8pt] coordinates {
    (30,0.10)
    (60,0.17)
    (120,0.36)
    (190,0.69)
    (280,1.02)
};
\addlegendentry{{\tiny Gradient descent step time}}

\end{axis}
\end{tikzpicture}
\caption{Computation time of task \textbf{T2} on Cora dataset}
\label{fig:cora_runtime}
\end{figure}
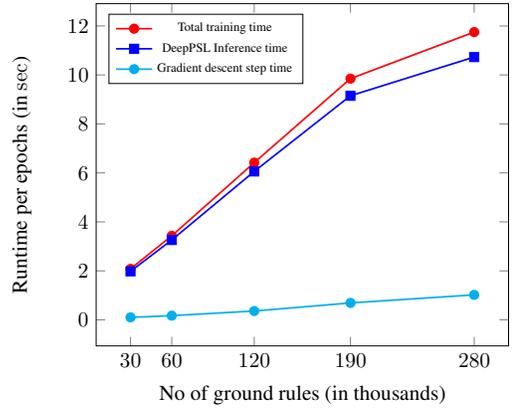

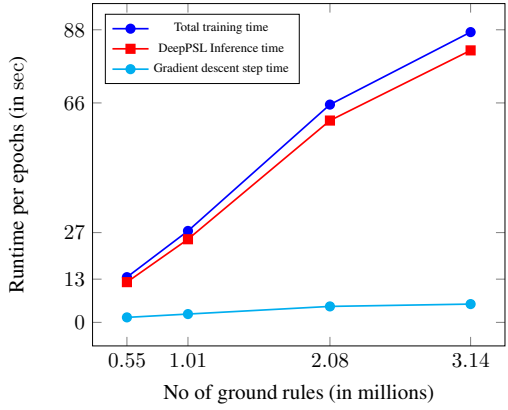
\begin{figure}[hbt]
\centering
\begin{tikzpicture}[scale=0.8]
\begin{axis}[
	legend style={at={(0.03,0.97)}, anchor=north west},
	xlabel={No of ground rules (in millions)},    
	ylabel={Runtime per epochs (in $\sec$)},    
	xtick={.55, 1.01, 2.08, 3.14},
	ytick={0, 13, 27, 66, 88},
	    ]
\addplot+[blue, mark options = {blue}, line width=0.8pt] coordinates {
    (0.55,13.6)
    (1.01,27.5)
    (2.08,65.5)
    (3.14,87.3)
};
\addlegendentry{{\tiny Total training time}}

\addplot+[red, mark options={red}, line width=0.8pt] coordinates {
    (0.55,12.1)
    (1.01,25)
    (2.08,60.7)
    (3.14,81.8)
};
\addlegendentry{{\tiny DeepPSL Inference time}}

\addplot+[cyan, mark options={cyan}, line width=0.8pt] coordinates {
    (0.55,1.5)
    (1.01,2.5)
    (2.08,4.8)
    (3.14,5.5)
};
\addlegendentry{{\tiny Gradient descent step time}}

\end{axis}
\end{tikzpicture}
\caption{Computation time of task \textbf{T3}}
\label{fig:runtime_er}
\end{figure}

We demonstrate experimentally that DeepPSL fully inherits the scalability of PSL, that is, they have the same Big O behavior. To this end, we perform run time and scalability analysis for semi-supervised classification task (\textbf{T2}) and entity resolution task (\textbf{T3}). We report average training and inference time (per epochs) of DeepPSL over 5 random training runs. For task \textbf{T2}, we sampled five sub-graphs from the citation network of Cora dataset and report their run time in Figure \ref{fig:cora_runtime} and Table \ref{tab:runtime_cora}. For task \textbf{T3}, we sampled four sub-graphs from the entity resolution dataset and report their run time in Figure \ref{fig:runtime_er} and Table \ref{tab:runtime_er}. For both tasks, the results indicate that the DeepPSL inference scales approximately linearly with number of ground rules.

The DeepPSL inference consists of two steps: The first step is a forward pass through neural network to evaluate neural predicates. The second step is standard PSL inference for determining inferred predicates, that is extremely efficient and scales linearly in number of ground rules. As discussed in Section 3.1 and \cite{bach2017hinge}, PSL inference easily scales to systems with millions of ground rules by leveraging the continuous nature of HL-MRFs that cast MAP inference as a convex optimization problem.

For example, in task \textbf{T2}, the neural predicates \pslpred{Neural} and \pslpred{Similar} are computed by a forward pass through the classifier and then LABEL is obtained by executing PSL inference. Since forward pass through the neural classifier takes negligible time as compared to PSL inference, the DeepPSL inference time will effectively be the same as PSL inference. Hence, DeepPSL inference scales linearly with number of ground rules.

The DeepPSL training algorithm, see Algorithm \ref{alg:opt}, executes DeepPSL inference step 
\[
\y_{t} = \argmin_{\y} \,\, \tilde{f} (\u, \w_{t    -1}, \y)
\]
and a gradient descent step  
\[
\begin{aligned} \w_t \mathrel{-}=  \eta  \nabla_{\w_t} \big[& \tilde{f}(\u, \w_t, \y_t - \alpha \nabla_{\y}L(\hat{\y},\y_t)) \\
& - \tilde{f}(\u,\w_t, \y_t) + \mu \Omega(\u, \w_t,\y_t)\big]
\end{aligned}
\]
for neural network training and rule weights learning. 

From Figure \ref{fig:cora_runtime} and Figure \ref{fig:runtime_er}, we note that DeepPSL time for every epoch also scales linearly with number of ground rules. It is evident from the experiments, see Table 4, that the run time of the gradient descent step is negligible as compared to DeepPSL inference step. Hence, from a computation perspective, the training of DeepPSL is equivalent to performing repeated DeepPSL inferences. The inference time for DeepPSL is approximately the same as PSL, and therefore, the training time for DeepPSL will be approximately equal to the time for repeated PSL inferences. Since PSL inference scales linearly with number of ground rules, DeepPSL training also approximately scales linearly with number of ground rules as observed in experiments, see Figure \ref{fig:cora_runtime} and Figure \ref{fig:runtime_er}.

\begin{table}[hbt]
    \caption{Run time statistics of task \textbf{T2} on Cora dataset}
    \label{tab:runtime_cora}
    \centering
\begin{threeparttable}
    \begin{tabular}{llllll}
    \toprule
    Nodes       & Edges & Ground rules & Inf\tnote{$\star$}  &  GDs\tnote{$\dagger$} & Total\tnote{$\ddagger$}  \\
    \midrule
    500       & 418 & 30352 & 1.98  &  0.10 & 2.08  \\
    1000       & 1476 & 69664 & 3.26  &  0.17 & 3.43  \\
    1500       & 3334 & 120176 & 6.06  &  0.36 & 6.42  \\
    2000       & 6572 & 190008 & 9.15  &  0.7 & 9.85  \\
    2708       & 10556 & 280476 & 10.73  &  1.02 & 11.75  \\
    \bottomrule
    \end{tabular}
    \begin{tablenotes}
    \item[$\star$] {\footnotesize DeepPSL Inference time per epochs (in $\sec$).}
    \item[$\dagger$] {\footnotesize Gradient descent step time per epochs (in $\sec$).}
    \item[$\ddagger$] {\footnotesize Total training time per epochs (in $\sec$).}
    \end{tablenotes}
\end{threeparttable}
\end{table}

\begin{table}[hbt]
    \caption{Run time statistics of task \textbf{T3}}
    \label{tab:runtime_er}
    \centering
\begin{threeparttable}
    \begin{tabular}{llll}
    \toprule
    Ground rules & Inf Time\tnote{$\star$}  &  GDs Time\tnote{$\dagger$} & Total Time\tnote{$\ddagger$}  \\
    \midrule
    551220 & 12.1  &  1.5 & 13.6  \\
    1011922 & 25  &  2.5 & 27.5  \\
    2083797 & 60.7  &  4.8 & 65.5  \\
    3143623 & 81.8  &  5.5 & 87.3  \\
    \bottomrule
    \end{tabular}
    \begin{tablenotes}
    \item[$\star$] {\footnotesize DeepPSL Inference time per epochs (in $\sec$).}
    \item[$\dagger$] {\footnotesize Gradient descent step time per epochs (in $\sec$).}
    \item[$\ddagger$] {\footnotesize Total training time per epochs (in $\sec$).}
    \end{tablenotes}
\end{threeparttable}
\end{table}

\section{DeepPSL Hyper-parameters Study}
We conduct experiments for studying effect of various hyperparameters on the convergence of DeepPSL. The following hyperparameters are considered for the study while training DeepPSL on task \textbf{T2} with Cora dataset: 
\begin{enumerate}[wide, label=\bf{\arabic*.}, labelwidth=!]
\item random initialization of neural network weights and rule weights,
\item regularization parameter $\mu$,
\item neural network weights's learning rate $\eta_{\omega}$,
\item neural network's hidden nodes and dropout rate.
\end{enumerate}

\subsection{Initialization of neural weights and rule weights}
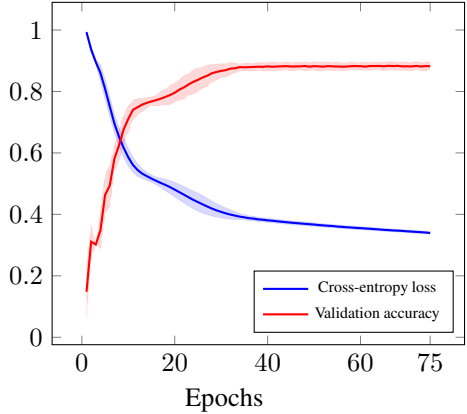
\begin{figure}[hbt]
\centering
\usepgfplotslibrary{fillbetween}
\usetikzlibrary{intersections}
\begin{tikzpicture}[scale=0.8]
\begin{axis}[
    legend style={at={(0.59,0.25)}, anchor=north west},
    xlabel={Epochs},
    xtick={0, 20, 40, 60, 75},
]
\addplot+[mark=none, line width=0.8pt] table[x=epochs, y=avg_loss, col sep=comma]{valNloss.csv};
\addlegendentry{{\tiny Cross-entropy loss}}
\addplot+[mark=none,line width=0.8pt] table[x=epochs, y=avg_val_acc, col sep=comma]{valNloss.csv};
\addlegendentry{{\tiny Validation accuracy}}
\addplot+[
    name path=upper,
    mark=none,
    blue,
    opacity=0.1,
    ] table[x=epochs, y=up_loss, col sep=comma]{valNloss.csv};

\addplot+[
    name path=lower,
    mark=none,
    blue,
    opacity=0.1,
    ] table[x=epochs, y=low_loss, col sep=comma]{valNloss.csv};

\addplot[
    blue,
    fill=blue,
    fill opacity=0.15,
    ] fill between[of=upper and lower];

\addplot+[
    name path=upper_val,
    mark=none,
    red,
    opacity=0.1,
    ] table[x=epochs, y=up_val_acc, col sep=comma]{valNloss.csv};

\addplot+[
    name path=lower_val,
    mark=none,
    red,
    opacity=0.1,
    ] table[x=epochs, y=low_val_acc, col sep=comma]{valNloss.csv};

\addplot[
    blue,
    fill=red,
    fill opacity=0.15,
    ] fill between[of=upper_val and lower_val];
\end{axis}
\end{tikzpicture}
\caption{Task \textbf{T2} on Split 2 of Cora dataset}
\label{fig:cora_s2}
\end{figure}

The DeepPSL rule weights are randomly initialized by drawing a sample from normal distribution $\mathcal{N}(1, 0.1)$ and the neural network weights are initialized using standard techniques. We train the DeepPSL system for task \textbf{T2} on Split 2 of the Cora dataset. We take average of cross-entropy loss and validation accuracy over 10 random runs with random weights initialization and random data splits, see Figure \ref{fig:cora_s2}. These experiments show that DeepPSL is quite robust to random initialization of neural and rule weights. 

\subsection{Regularization parameter}
In this experiment, we varied the regularization hyperparameter $\mu$ among the set $\{0.04, 0.12, 0.2\}$ and found that DeepPSL trains effectively and reaches the desired performance across a broad range of the regularization parameter. As shown in Figure \ref{fig:cora_reg}, the validation accuracy and cross-entropy loss indicate that DeepPSL does not become trapped in local minima.  
\begin{figure}[hbt]
\centering
\begin{tikzpicture}[scale=0.8]
\begin{axis}[
    legend style={at={(0.22,0.97)},legend columns=2, anchor=north west},
    xlabel={Epochs},
    xtick={0, 20, 40, 60, 80, 100},
]
\addplot+[mark=none, red, line width=0.8pt] table[x=epochs, y=l1_reg_0.04, col sep=comma]{hyper_reg.csv};
\addlegendentry{{\tiny CE loss, $\mu = 0.04$}}
\addplot+[mark=none, blue, line width=0.8pt] table[x=epochs, y=v1_reg_0.04, col sep=comma]{hyper_reg.csv};
\addlegendentry{{\tiny Val acc, $\mu= 0.04$}}
\addplot+[mark=none, red!30!green, line width=0.8pt] table[x=epochs, y=l1_reg_0.12, col sep=comma]{hyper_reg.csv};
\addlegendentry{{\tiny CE loss, $\mu= 0.12$}}
\addplot+[mark=none, green, line width=0.8pt] table[x=epochs, y=v1_reg_0.12, col sep=comma]{hyper_reg.csv};
\addlegendentry{{\tiny Val acc, $\mu= 0.12$}}
\addplot+[mark=none, black, line width=0.8pt] table[x=epochs, y=l1_reg_0.2, col sep=comma]{hyper_reg.csv};
\addlegendentry{{\tiny CE loss, $\mu= 0.2$}}
\addplot+[mark=none, cyan, line width=0.8pt] table[x=epochs, y=v1_reg_0.2, col sep=comma]{hyper_reg.csv};
\addlegendentry{{\tiny Val acc, $\mu= 0.2$}}
\end{axis}
\end{tikzpicture}
\caption{Task \textbf{T2} on Split 1 of Cora dataset}
\label{fig:cora_reg}
\end{figure}

\subsection{Learning rate of neural network weights}
In this experiment, we varied the learning rate of neural network weights $\eta_{\omega} \in \{ 0.004, 0.012, 0.1 \}$ to examine the robustness of DeepPSL training across different learning rates. We found that while small learning rates slow down training, large learning rates can lead to oscillations, as seen in Figure \ref{fig:cora_lr}. However, the validation accuracy and cross-entropy loss, see Figure \ref{fig:cora_lr}, indicate that DeepPSL training is robust for a wide range of learning rate parameters.  
 
\begin{figure}[htb]
\centering
\begin{tikzpicture}[scale=0.8]
\begin{axis}[
    legend style={at={(0.14,0.97)},legend columns=2, anchor=north west},
    xlabel={Epochs},
    xtick={0, 20, 40, 60, 80, 100},
]
\addplot+[mark=none, red, line width=0.8pt] table[x=epochs, y=l1_l2_0.1, col sep=comma]{hyper_lr.csv};
\addlegendentry{{\tiny CE loss, $\eta_{\omega} = 0.1$}}
\addplot+[mark=none, blue, line width=0.8pt] table[x=epochs, y=v1_l2_0.1, col sep=comma]{hyper_lr.csv};
\addlegendentry{{\tiny Val acc, $\eta_{\omega} = 0.1$}}
\addplot+[mark=none, red!30!green, line width=0.8pt] table[x=epochs, y=l1_l2_0.02, col sep=comma]{hyper_lr.csv};
\addlegendentry{{\tiny CE loss, $\eta_{\omega} = 0.02$}}
\addplot+[mark=none, green, line width=0.8pt] table[x=epochs, y=v1_l2_0.02, col sep=comma]{hyper_lr.csv};
\addlegendentry{{\tiny Val acc, $\eta_{\omega} = 0.02$}}
\addplot+[mark=none, black, line width=0.8pt] table[x=epochs, y=l1_l2_0.002, col sep=comma]{hyper_lr.csv};
\addlegendentry{{\tiny CE loss, $\eta_{\omega} = 0.002$}}
\addplot+[mark=none, cyan, line width=0.8pt] table[x=epochs, y=v1_l2_0.002, col sep=comma]{hyper_lr.csv};
\addlegendentry{{\tiny Val acc, $\eta_{\omega} = 0.002$}}
\end{axis}
\end{tikzpicture}
\caption{Task \textbf{T2} on Split 1 of Cora dataset}
\label{fig:cora_lr}
\end{figure}

\subsection{Neural network hyperparameters}
We varied the number of hidden nodes and dropout rate of the neural classifier and found that DeepPSL training is relatively robust to these hyperparameters, as shown in Table \ref{tab:neural_hyp}. However, the performance of DeepPSL decreases with fewer hidden nodes and higher dropout rates, indicating that this combination of hyperparameters is not sufficient to capture the underlying structure of the data.
\begin{table}[htb]
    \caption{Classification accuracy of DeepPSL on task \textbf{T2} (Split 1 of Cora dataset)}
    \label{tab:neural_hyp}
    \centering
\begin{threeparttable}
\begin{tabular}{|l||*{3}{c|}@{\hspace{-0.2ex}}}\hline
\multicolumn{1}{|@{}l||}{\backslashbox[0pt][l]{Drp\tnote{$\star$}}{Nds\tnote{$\dagger$}}}
&\makebox[3em]{\textbf{8}}&\makebox[3em]{\textbf{16}}&\makebox[3em]{\textbf{24}}\\\hline\hline
\textbf{0}  & $80.12 \pm 0.70$ & $80.42 \pm 0.49$ & $80.64 \pm 0.35$ \\\hline
\textbf{0.2}& $79.97 \pm 0.60$ & $81.85 \pm 0.43$ & $81.76 \pm 0.45$\\\hline
\textbf{0.4}& $74.23 \pm 0.78$ & $79.16 \pm 0.57$ & $80.47 \pm 0.70$\\\hline
\end{tabular}
    \begin{tablenotes}
    \item[$\star$] {\footnotesize Dropout rate in neural classifier.}
    \item[$\dagger$] {\footnotesize Hidden node of neural classifier.}
    \end{tablenotes}
\end{threeparttable}
\end{table}

\section{Toy example for DeepPSL} 
To explain DeepPSL in a simple and intuitive fashion, we discuss a toy version of the entity resolution problem (explained in Section \ref{exp_ER}) to first elucidate the core concepts of PSL and then introduce the DeepPSL framework. In the toy problem, the goal is to identify the duplicate author references based on similarity of  author names. Using the following PSL rules, we express the knowledge that two authors are same if they have similar names and that there is a transitive relation.
 
\begin{figure}[htpb] 
    \centering
    \noindent\fbox{%
        \begin{minipage}{0.95\hsize}
            \begin{scriptsize}
            \begin{flalign*}
                \hspace{0.2cm} & \textit{\textbf{\# Entity resolution rules}}&& \\
                \theta_{1}:& \pslpred{AuthorName}(\pslarg{A1}, \pslarg{N1}) \psland \pslpred{AuthorName}(\pslarg{A2}, \pslarg{N2}) \psland (\pslarg{A1}\pslneg = \pslarg{A2}) \psland\\ 
			   & \pslpred{SimName}(\pslarg{N1}, \pslarg{N2}) \pslthen \pslpred{SameAuthor} (\pslarg{A1}, \pslarg{A2}) \\
                \theta_{2}:& \pslpred{SameAuthor}(\pslarg{A1}, \pslarg{A2}) \psland \pslpred{SameAuthor}(\pslarg{A2}, \pslarg{A3}) \psland (\pslarg{A1}\pslneg = \pslarg{A2}) \psland \\  
			   & (\pslarg{A2}\pslneg = \pslarg{A3}) \psland (\pslarg{A1}\pslneg = \pslarg{A3}) \pslthen \pslpred{SameAuthor} (\pslarg{A1}, \pslarg{A3}) \\
            \end{flalign*}
            \end{scriptsize}
        \end{minipage}
    \hspace{-0.4cm}}
    \caption{DeepPSL rule set for entity resolution}
\end{figure}

Here, $\pslpred{SameAuthor}$ is an inferred predicate and, $\pslpred{AuthorName}, \pslpred{SimName}$ are observed predicates. Consider the scenario where there are 3 authors mention with names, 
\[
\mathcal{L} = \{\text{``Sheldon Cooper"}, \text{``Cooper Sheldon"}, \text{``Sheldon C"}\}
\] 
with their corresponding \pslarg{IDs} in 
\[ \mathcal{D} = \{\pslarg{AID1}, \pslarg{AID2}, \pslarg{AID3}\}\] respectively. The base of ground atoms is given by
\begin{align}
\mathcal{A} = \left\{\hspace{-0.13cm} 
\begin{array}{l} 
x_{ij} = \pslpred{AuthorName}(i, j)\; \text{for}\;  i \in \mathcal{D} \; \text{and}\; j \in \mathcal{L},\\
\bar{x}_{ij} = \pslpred{SimName}(i, j) \phantom{,j)}\; \text{for} \; i , j \in \mathcal{L},\\
y_{ij} = \pslpred{SameAuthor}(i, j) \phantom{,} \; \text{for} \; i, j \in \mathcal{D} \; \text{and} \; i \neq j 
\end{array}\hspace{-0.13cm}\right\} 
\end{align} 
The above rules will induce to the following ground rules:
\begin{align} 
\theta_1: \; \;x_{ij} \; \& \; x_{kl}\;& \& \; \bar{x}_{jl}\; \rightarrow\; y_{ik} \; \nonumber \\ &\text{for} \; i,k \in \mathcal{D},\; i \neq k, \;  \text{and} \; j,l \in \mathcal{L}\\
\theta_2: \; \; y_{ij} \; \& \; y_{jk}\;& \rightarrow\; y_{ik} \;\phantom{\; \& \; \bar{x}_{jl}}  \nonumber \\ &\text{for} \; i,j,k \in \mathcal{D}\; \text{and}\; i \neq j \neq k
\end{align} 
which is equivalent to the following:
\begin{align}\label{eq:dis_exp} 
\theta_1: \; !x_{ij} \; \| \; !x_{kl} \; \| \; !\bar{x}_{jl} \; \| \; y_{ik}, \qquad 
\theta_2: \; !y_{ij} \; \| \; !y_{jk} \; \| \; y_{ik}
\end{align} 
The disjunction expression \eqref{eq:dis_exp} is converted to a soft logic using Lukasiewicz t-norm and its corresponding co-norm as the relaxation of the logical AND and OR respectively. Additionally, negation of an atom, $``a"$ is considered as $``1-a"$. Therefore, the satisfaction of the rules \eqref{eq:dis_exp} are determined by the following expressions:
\begin{equation}\label{eq:rule_sat}
\begin{aligned}
&\min \left\{\left(1-x_{ij}\right) + \left(1-x_{kl}\right) + \left(1-\bar{x}_{jl}\right) + y_{ik}, 1\right\},\\ 
&\min \left\{\left(1-y_{ij}\right) + \left(1-y_{jk}\right) + y_{ik}, 1\right\}.
\end{aligned}
\end{equation}
The rule satisfaction \eqref{eq:rule_sat} are rewritten as the distance to satisfaction leading to the following hinge-loss potentials:
\begin{align*} 
\begin{cases}
\phi_{ijkl} = \left(\max\{x_{ij} + x_{kl} + \bar{x}_{jl} - y_{ik} - 2, 0\}\right)^p\\
\bar{\phi}_{ijk} = \left(\max\{y_{ij} + y_{jk} - y_{ik} - 1, 0\}\right)^p
\end{cases} 
\end{align*}  
for $p \in \{1, 2\}.$
With $\x := \left(x_{ij}, \bar{x}_{ij}\right)$, $\y := \left(y_{ij}\right)$, $\t:=\left(\theta_1, \theta_2\right)$, the energy function is defined as
\begin{align}
f_{\t} (\x, \y) = \theta_1 \sum_{\substack{i,k \in \mathcal{D} \\ i \neq k \\ j,l \in \mathcal{L}}} \phi_{ijkl} + \theta_2 \sum_{\substack{i,j,k \in \mathcal{D}\\ i\neq j\neq k}} \bar{\phi}_{ijk}
\end{align} 
The inferred predicates $\y$ are obtained by solving a convex PSL inference optimization (4).

For DeepPSL, the string similarity values need not be supplied beforehand. A siamese network can be plugged into the SimName predicate which computes the similarity of strings. We pass the string pairs to the neural network, get the similarity values and feed them as groundings for SimName predicate. Once the groundings are obtained, similar to PSL we minimize the total hinge loss to obtain the inferred values.

 
\end{document}